\documentclass{aa}  

\usepackage{graphicx}

\usepackage{txfonts}
\usepackage{newtxtext,newtxmath}
\usepackage{amsmath}
\usepackage[autostyle]{csquotes}
\usepackage{subfig}
\usepackage{xcolor}
\usepackage{hyperref}
\usepackage{comment}
\usepackage{multirow}

\begin{document}

   \title{Revealing the intricacies of radio galaxies and filaments in the merging galaxy cluster Abell 2255}
   \subtitle{I. Insights from deep LOFAR-VLBI sub-arcsecond resolution images}
   
   \titlerunning{High-resolution radio filaments in Abell 2255}
   
   \authorrunning{De Rubeis et al.}

   \author{E. De Rubeis\inst{1,2},
            M. Bondi\inst{2},
            A. Botteon\inst{2},
            R. J. van Weeren\inst{3},
            J. M. G. H. J. de Jong\inst{3,4},
            L. Rudnick\inst{5},
            G. Brunetti\inst{2},
            K. Rajpurohit\inst{6},
            C. Gheller\inst{2},
            H. J. A. R{\"o}ttgering\inst{3}
          }

   \institute{Dipartimento di Fisica e Astronomia, Universit\`a di Bologna, via Gobetti 93/2, I-40129 Bologna, Italy\\
              \email{emanuele.derubeis2@unibo.it}
              \and
    INAF - Istituto di Radioastronomia di Bologna, Via Gobetti 101, I-40129 Bologna, Italy
    \and
    Leiden Observatory, Leiden University, PO Box 9513, NL-2300 RA Leiden, The Netherlands
    \and
    ASTRON, The Netherlands Institute for Radio Astronomy, Postbus 2, 7990 AA Dwingeloo, The Netherlands
    \and
    Minnesota Institute for Astrophysics, University of Minnesota, 116 Church St. SE, Minneapolis, MN 55455, USA
    \and
    Center for Astrophysics | Harvard \& Smithsonian, 60 Garden St., Cambridge, MA 02138, USA}

   \date{Received XXX; accepted YYY}
 
  \abstract 
   {High sensitivity of modern interferometers is revealing a plethora of filaments surrounding radio galaxies, especially in galaxy cluster environments. The morphology and spectral characteristics of these thin structures require the combination of high-resolution and low frequency observations, which is best obtained using the LOw Frequency ARray (LOFAR) international stations.}
   {In this paper, we aim to detect and characterize non-thermal filaments observed close or as part of the radio galaxies in Abell 2255 using deep, LOFAR-VLBI observations at 144 MHz. These structures can be used to disentangle possible scenarios for the origin of the non-thermal filaments and connection to the motion of the host galaxy within the dense and turbulent intracluster medium (ICM), and consequent interaction between the ICM and radio jets.}
   {Combining multiple observations, we produced the deepest images ever obtained with LOFAR-VLBI targeting a galaxy cluster, using 56 hours of observations, reaching $0.3-0.5"$ resolution. We detailed throughout the paper the calibration and imaging strategy for the different targets, as well as the multitude of morphological features discovered.}
   {Thanks to the high-sensitivity of LOFAR-VLBI, we revealed unprecedented details for the main cluster radio galaxies, recovering in most cases also their more extended structure observed only at such low frequencies. In particular, we focused on the Original Tailed Radio Galaxy (Original TRG) where we distinguished many filaments constituting its tail with varying lengths ($80-110$ kpc) and widths ($3-10$ kpc). The final radio images showcase the potential of deep, high-resolution observations for galaxy clusters. With such approach, we enabled the study of these thin, elongated radio filaments: after being discovered, these filaments now require spectral studies to determine their formation mechanisms.}
   {}


   \maketitle
%

\section{Introduction}
\label{sec:introduction}
In galaxy clusters (GC), radio galaxies interact with the dense ambient intracluster medium (ICM), typically showing distorted radio morphologies. The synchrotron emitting jets ejected from the host galaxy can be bent by ram pressure effects, while buoyancy effects can cause them to move towards the edge of the cluster. Based on the bending angle of the two jets, the radio morphology varies from wide-angle tails (WAT) to narrow-angle tails (NAT) to head-tails (HT), where the jets are bent in a common direction following the wake of the host galaxy with the latter representing the head~\citep{1980miley,2020hardcastle}. High sensitivity observations, especially with modern interferometers, are revealing the presence of a wealth of filaments both in the tail and in the surroundings of radio galaxies, particularly for the ones residing in a cluster or group environment~\citep[e.g., ][]{hardcastle2019, lal2020, ramatsoku2020, brienza2021, condon2021, rudnick2022, candini2023, koribalski2024}. Simulations show that extended bundles of magnetic fields, which would be observed as filaments, are ubiquitous in turbulent magnetohydrodynamical (MHD) flows where their lengths reflect the local driving scales of the turbulence~\citep{porter2015}. Filamentary structures present new opportunities for studying the physical processes in the ICM, including their magnetic structures and the propagation of cosmic rays. They can serve as probes of shear motions, revealing the driving and dissipation scales of dynamic structures as the cluster evolves, and their widths can also provide information about the resistivity scales in the plasma. Filaments can determine an alternative site for cosmic-ray acceleration in clusters: in fact, according to~\citet{bell2019}, repeated encounters with weak shocks in magnetic flux tubes in the backflows of radio galaxies could accelerate electrons to extremely high energies. Ultimately, the seed electrons upon which the re-acceleration operates on cluster-scales likely come from current or past active galactic nuclei (AGN) activity: given that particle acceleration via shocks and turbulence is an inefficient process~\citep[for a complete review see][]{2014brunetti}, these seed electrons are thought to be essential for the formation of diffuse emission in the form of radio halos and relics~\citep{vazza2024}. Thus, beyond simple ram-pressure deflection, we must explore a wider range of the mutual interactions between ICM and radio galaxies to unveil the physical mechanisms responsible for the complex morphology of these objects~\citep{rudnick2022}. Observing distorted filaments and substructure of nearby radio galaxies gives us a unique opportunity to explore the nature and origin of these features, in particular the re-energization processes and the role of internal and external magnetic fields. From the literature, these features are known to have steep radio spectra ($\alpha>1.3$~\footnote{We use the convention $S(\nu) \propto \nu^{-\alpha}$ for synchrotron spectrum, with $\alpha > 0$}), with signs of steepening along the filaments' length and width~\citep{rudnick2022,brienza2025}, projected lengths ranging between 10s-100s kpc, and widths from a few kpc down to $350-500$ pc~\citep[as observed in Perseus cluster,][]{vanweeren2024}. These characteristics underline the necessity of sensitive, low-frequency and high-resolution observations to detect and disentangle such filaments.
\par The International LOw Frequency ARray (LOFAR) Telescope~\citep[ILT,][]{vanhaarlem2013} represents an excellent candidate for this scope. Having baselines extending to almost 2,000 km, this interferometer provides a unique resolution down to 0.3" at the high-band antenna (HBA) frequencies of 144 MHz ~\citep[][]{morabito2022}, making it ideal for targeting filaments in tails of radio galaxies and within the ICM. Recently, the calibration and imaging procedure has been standardized with the development of a LOFAR Very Long Baseline Interferometry (VLBI) pipeline~\citep{morabito2022}, capable of calibrating HBA data using full ILT array and imaging single target sources down to sub-arcsecond resolution at 144 MHz. This pipeline has been recently upgraded to a new version which implements the Common Workflow Language (CWL), becoming the new standard for LOFAR-VLBI data~\footnote{\url{https://git.astron.nl/RD/VLBI-cwl}} (van der Wild \textit{in preparation}). In the past years, this pipeline has been successfully used for targeting galaxy clusters and groups~\citep[e.g.,][]{timmerman2022, cordun2023, timmerman2024, ubertosi2024, vanweeren2024, pasini2025}.
\par
\begin{figure*}[ht!]
\centering
\includegraphics[width=0.8\hsize]{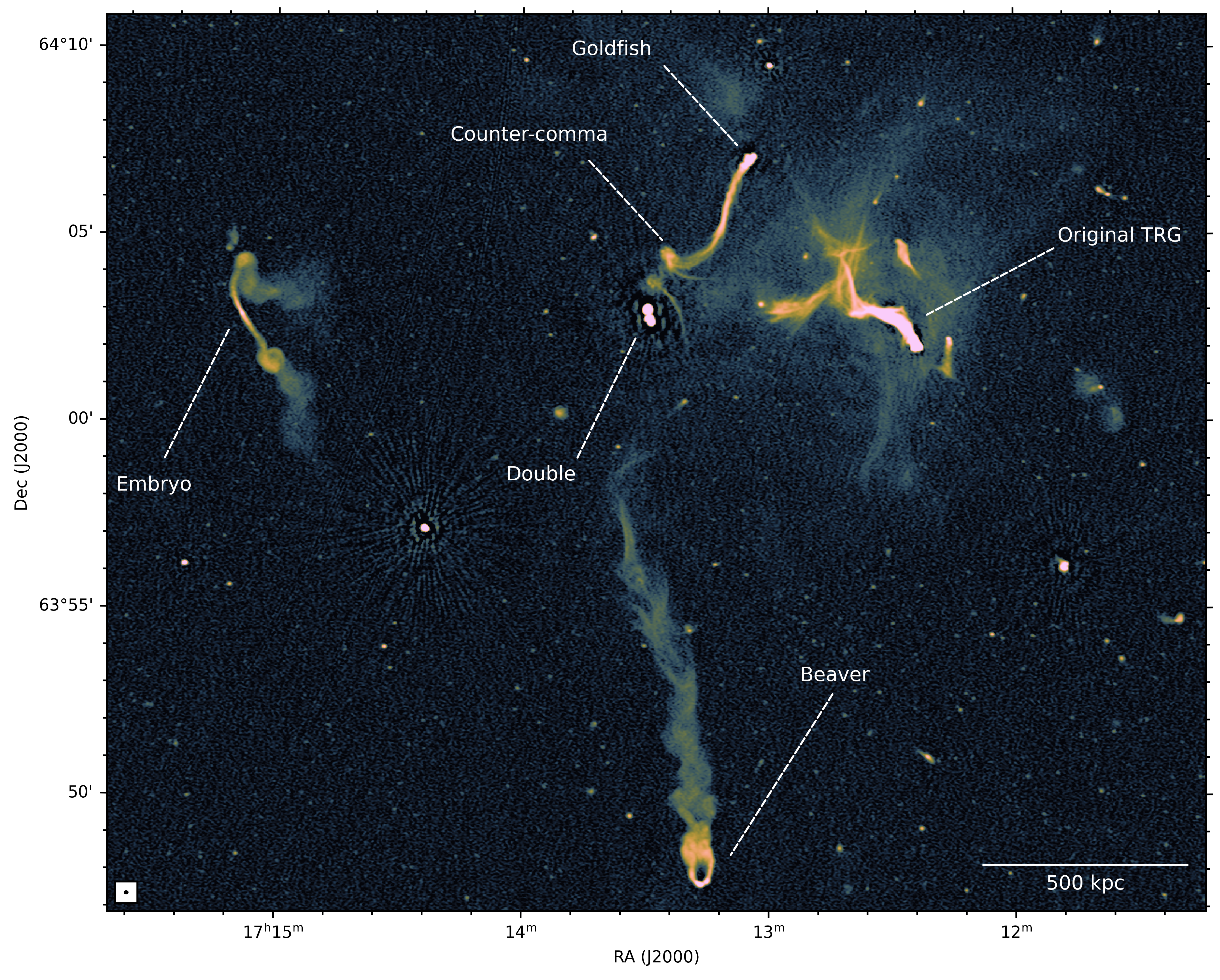}
\caption{Zoom-in at Abell 2255's center at 145 MHz, resolution of $4.7" \times 3.5"$, adapted from~\citet{botteon2022}. The main cluster member radio galaxies, focus of this paper, are highlighted in white and are labeled as in~\citet{harris1980} and~\citet{botteon2020}. The restoring beam size is shown in the bottom-left corner.}
\label{fig:lofar_6aseclabel}
\end{figure*}

 In this paper, we used the ILT to observe the radio galaxies and filaments in the galaxy cluster Abell 2255 (hereafter A2255) at sub-arcsecond resolution. This is a merging galaxy cluster ($z = 0.0806$) showing morphologically complex radio structures on multiple scales~\citep[see][and references therein for a more detailed description]{botteon2020,botteon2022}. Several optical and infrared studies reported a total number of confirmed member galaxies that range from 300 to 500 (depending on the radius selection criteria), using both spectroscopic and photometric techniques~\citep{yuan2003, shim2011, tyler2014, golovich2019}.~\citet{miller2003} focused in particular on the fraction of AGN: using Karl G. Jansky Very Large Array (VLA) data at 1.4 GHz, they found that A2255 has an abnormal abundance of radio galaxies compared to other clusters, which is possibly motivated by the cluster's dynamical state, even if this is not yet clear from optical observations~\citep{burns1995, yuan2003}. Radio galaxies in A2255 have been widely observed in the past~\citep[e.g.,][]{harris1980, burns1995, miller2003, govoni2006, pizzo2009, pizzo2011, ternidegregory2017, botteon2020,botteon2022}. Four tailed radio galaxies reside in the cluster: following the nomenclature introduced by~\citet{harris1980}, there are three NAT sources, namely the Original Tailed Radio Galaxy (Original TRG), the Goldfish, and the Beaver, and a WAT, the Embryo. Together with these, there is also a Fanaroff-Riley II~\citep[FRII,][]{fanaroff1974} radio galaxy located, in projection, close to the cluster center called the Double. Observations for these sources were performed by~\citet{govoni2006} at high frequencies, using VLA in C- and X-band providing resolution of $2"$; they also studied the fractional polarization of these sources, finding mean values around $14\%$ for the NATs. The Goldfish has been observed also with VLA at 15 GHz, reaching an angular resolution of $0.47" \times 0.44"$~\citep{ternidegregory2017}, revealing strong projection effects impacting its morphology. Together with observations of cluster radio galaxies,~\citet{govoni2005} discovered the presence of strongly polarized filaments ($20-40\%$ of fractional polarization) in the cluster radio halo at 1.4 GHz, with rectangular shape and regions of ordered magnetic field of $\sim400$ kpc in size. They were reported also subsequently using the Westerbork Synthesis Radio Telescope (WSRT) by~\citet{pizzo2009} and~\citet{pizzo2011}. Recently, thanks to deep LOFAR observations (75 hours at 145 MHz, rms noise $43~\rm{\mu Jy~beam^{-1}}$, resolution $4.7" \times 3.5"$, Fig.~\ref{fig:lofar_6aseclabel}; 72 hours at 49 MHz, rms noise $730~\rm{\mu Jy~beam^{-1}}$, resolution $11.5" \times 8.2"$),~\citet{botteon2020,botteon2022} detected the presence of additional filamentary and distorted structures on different scales with a very steep spectrum ($\alpha > 2$) embedded in the radio halo, whose association with optical counterparts of the cluster radio galaxies is not trivial, considering also that these filaments are detectable only at LOFAR frequencies.
\par In this work, we aim to observe the tails and filaments that populate the cluster environment using the high-resolution, low-frequencies capabilities of the ILT. We produced radio maps at 144 MHz at sub-arcsecond resolution, targeting the brightest cluster member radio galaxies: using 56 hours of observations, this makes the deepest LOFAR-VLBI images ever targeting a galaxy cluster. Thanks to the high-resolution and sensitivity of LOFAR-VLBI we detected, for the first time, a multitude of sub-structures related to these radio galaxies, mostly filamentary, observable only at such low frequencies. Particular attention was given to the Original TRG, which showed the presence of filaments related to its tail extending for almost 300 kpc presuming a complex origin scenario involving the turbulent environment which surrounds it. In this paper, we focus on the methodology used to produce the deep LOFAR-VLBI images and on a morphological study of the Original TRG. In an upcoming paper, this analysis will be combined with high-resolution spectral index maps resulting from the combination of LOFAR-VLBI data with upgraded Giant Metrewave Radio Telescope (uGMRT) and the VLA at higher frequencies.
\par In this paper, we assume a flat $\rm{\Lambda}CDM$ cosmology, with $ H_{0} = 70~\rm{km~s^{-1}~Mpc^{-1}}$, $\Omega_{m} = 0.3$, and $\Omega_{\Lambda} = 0.7$. At the redshift of Abell 2255, $1"$ corresponds to a linear scale of 1.512 kpc.

\begin{table*}[ht!]
\centering
\caption{List of the LOFAR observations used in this paper.}
\label{tab:data}
\begin{tabular*}{0.6\textwidth}{c@{\extracolsep{\fill}}ccc}
\hline\hline
Obs. ID & Obs. date & Stations (International) & $\sigma_{\rm{rms},6"}$ \\
 & [YYYY-MM-DD] & & [$\rm{\mu Jy~beam^{-1}}$] \\
\hline\hline
L720378 & 2019-06-07 & 51 (13) & 89 \\
L725454 & 2019-06-22 & 51 (13) & 90 \\
L726708 & 2019-06-28 & 51 (13) & 85 \\
L727110 & 2019-07-03 & 50 (12) & 92 \\
L733077 & 2019-08-09 & 48 (13) & 97 \\
L747613 & 2019-09-28 & 47 (11) & 99 \\
L751366 & 2019-10-04 & 50 (12) & 93 \\
\hline\hline
\end{tabular*}
\tablefoot{Column 1: ID of the observation. Column 2: date of the observation. Column 3: total number of stations (number of IS in brackets). Column 4: rms noise of the map output from \texttt{ddfpipeline} using only Dutch stations.}
\end{table*}

\section{Data calibration and imaging}
\label{sec:data_calibration}
A2255 was first observed for 72 hours by LOFAR in the period June to November 2019 (project LC12\_017, P.I. R.J. van Weeren). The cluster is located at about 5 degrees from the Euclid Deep Field North (EDFN) and this made possible to observe the two targets simultaneously using two different LOFAR beams~\citep{bondi2024}. After these first observations, A2255 was again observed for an additional 226 hours as the secondary target during the very deep EDFN observations carried out in the period 2022-2023 (project LT16\_005, P.I. P.N Best). For this paper, we selected only observations coming from the first set of 72 hours, because the latest ones are still in the phase of processing. The 72 hours were split in 9 runs, and each 8 hour observation was calibrated individually.
\par After describing the preliminary steps involved in the calibration of the Dutch stations, for both direction-independent (DIE) and direction-dependent (DDE) effects, we detail the procedures on the international stations (IS): for their calibration strategy, we follow the procedures extensively described in~\citet{morabito2022}, with different tailored adjustments, especially in the imaging step, required by the high-complexity radio morphology of the field containing A2255, with the combination of extended and structured radio galaxies overlapping the cluster radio halo.

\subsection{Initial Dutch stations calibration}
\label{sec:initial_cal}
The first step provides calibration of the visibilities following the standard procedure used for the LOFAR Two-metre Sky Survey~\citep[LoTSS,][]{shimwell2017,shimwell2019,shimwell2022}. After downloading data from the LOFAR Long-Term Archive (LTA), both for primary calibrator and target, for each observation, we used \texttt{Prefactor}~\citep{degasperin2019} solutions from~\citet{botteon2020,botteon2022} on the primary calibrator, which is 3C48 for all the observations. This pipeline corrects for the polarization alignment, clock delays between stations, and bandpass, for both Dutch and international stations. \texttt{Prefactor} is used also for the target field (A2255) to correct for DIE on the Dutch stations. After applying the solutions found for the primary calibrator to the target, it calibrates the phases against a sky model from the TIFR GMRT Sky Survey~\citep[TGSS,][]{intema2017}. 
\par We then further processed our data to correct for DDE using \texttt{ddf-pipeline}~\citep{shimwell2017, tasse2021}, which uses \texttt{killMS}~\citep{tasse2014,smirnov2015} for direction-dependent calibration and \texttt{DDFacet}~\citep{tasse2018} to apply direction-dependent solutions during imaging. The output image of the target field at $6"$ resolution, corrected for both DIE and DDE on the Dutch stations, has been used to assess the quality of the observation.

\subsection{International stations calibration}
\label{sec:is_calibration}
Once the Dutch array is fully calibrated, we proceeded with the calibration of the IS. Following the strategy described in~\citet{morabito2022}, solutions from the Dutch array corrected for both DIE and DDE were transferred to the original measurement sets including also the IS. Then, after checking the contribution of bright, off-axis sources (so-called A-team sources, that were sufficiently far to avoid contaminating the target field), datasets were concatenated into sub-bands of 1.95 MHz each. These sub-bands were phase-shifted towards the direction of an in-field calibrator for the calibration of the IS, required for correcting direction-independent dispersive delays, and concatenated in frequency into a single measurement set.
\par For our field, we searched for the best available in-field calibrator using the Long Baseline Calibrator Survey~\citep[LBCS,][]{jackson2016,jackson2022}. The selected source was 4C $+64.21$~\citep[RA: $\rm{17^h 19^m 59^s}$, Dec: $\rm{64^{\circ}04'37"}$,][]{pilkington1965}, a powerful radio source, compact on arcsec-scale, with flux density of $\sim 4.87~\rm{Jy}$ at 144 MHz. Given that it shows the presence of two components at $0.3"$ resolution, and it was observed at multiple frequencies throughout the radio spectrum, we built a sky model distributing the flux density among two Gaussian components, with the same flux ratio as the one in the full resolution map. To better reproduce the spectral index for the calibrator, we used flux density measures at multiple radio frequencies, following~\citet{dejong2024}: in particular, the 8C survey at 38 MHz~\citep{hales1995}, the VLA Low-frequency Sky Survey at 74 MHz~\citep[VLSS,][]{cohen2007}, the 6C survey at 151 MHz~\citep{hales1990}, the Westerbork Northern Sky Survey at 325 MHz~\citep[WENSS,][]{rengelink1997}, the Texas Survey at 365 MHz~\citep{douglas1996}, and the NRAO VLA Sky Survey~\citep[NVSS,][]{condon1998} at 1.4 GHz. With flux densities and frequencies we fitted a second-order logarithmic polynomial
\begin{equation}
\log{S(\nu)} = \log{S_{0}} + c_{0}\log{\Bigl(\frac{\nu}{\nu_{0}}\Bigr)}+c_{1}\log{\Bigl(\frac{\nu}{\nu_{0}}\Bigr)^2}~.
\end{equation}
Using $\nu_0 = 144~\rm{MHz}$ as the reference frequency, we ended up with $c_{0}=-0.013$ and $c_{1}=-0.90$. These parameters, together with the position of the Gaussian sub-components, made the initial sky model that we used for the in-field calibrator.
\par
\begin{figure*}[ht!]
\centering
\includegraphics[width=\hsize]{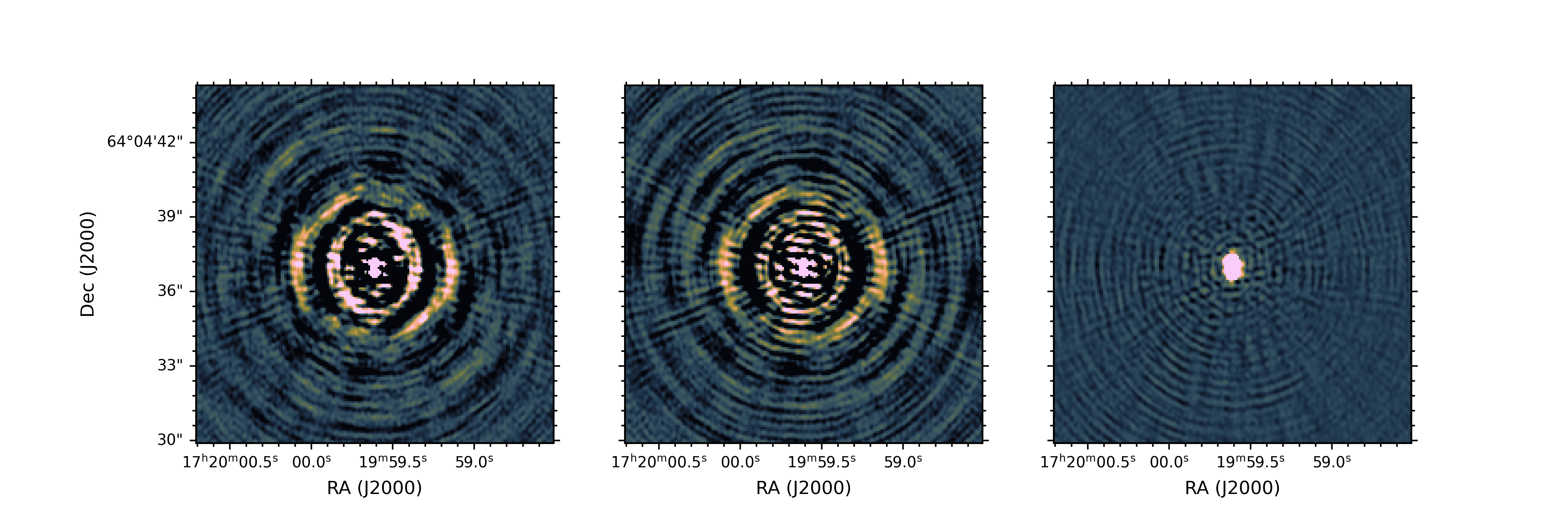}
\caption{Self-calibration results on the delay calibrator 4C $+64.21$ for one observing run. From left to right: self-calibration cycle 0 (sole imaging), cycle 1 (only \texttt{scalarphasediff} and \texttt{scalarphase} solutions), cycle 19 (final image, solving also for \texttt{scalarcomplexgain}).}
\label{fig:infield}
\end{figure*}

 We used \texttt{facetselfcal}~\citep{vanweeren2021}, which uses the \texttt{Default Preprocessing Pipeline}~\citep[DP3,][]{vandiepen2018,dijkema2023} and \texttt{WSClean}~\citep{2014offringa, offringa2017} to perform self-calibration. It allows for correcting phases and amplitudes minimizing the differences between the data and the sky model, with the latter that gets updated at each self-calibration cycle. We followed the procedure adopted by~\citet{dejong2024} with some minor adjustments. In particular, during the self-calibration cycles we solved for:
\begin{enumerate}
\item \texttt{scalarphasediff}: solution interval 128 s, frequency smoothness 4.0 MHz;
\item \texttt{scalarphase}: 32 s, frequency smoothness 1.0 MHz;
\item \texttt{scalarcomplexgain}: 53 min, frequency smoothness 4.0 MHz, one solution each 10 channels.
\end{enumerate}
We ignored all the baselines shorter than $40,000\lambda$, corresponding to an angular scale of about $5"$ at 144 MHz, using the parameter \texttt{-{}-uvmin}, to prevent possible incompleteness in the sky model. We phased up LOFAR's core stations to form a large, virtual, narrow field of view (FoV) station, to reduce the interference from unrelated nearby radio sources especially at short baselines. Through \texttt{facetselfcal}, we also averaged the input data down to 32s in time and 488 kHz in frequency and used a Briggs weighting with robustness of $-1.5$ for the imaging~\citep{1995briggs}. At the end of the self-calibration routine on the in-field calibrator, one Hierarchical Data Format 5 (\texttt{h5parm}) solution file for each calibration step and one containing all the solutions merged, with phases and amplitudes corrections, are produced. We performed 20 self-calibration cycles providing the best solutions for visibilities (Fig.~\ref{fig:infield}). Among the 9 available observations, two (namely L728074 and L746864) have been discarded because of the significantly higher image noise once including the IS (between $30-40\%$ with respect to the other runs). We ended up, then, with 7 runs (meaning 56 hours of observing time on target), that have been used for the long baselines calibration and imaging. The main characteristics of the selected observations are listed in Tab.~\ref{tab:data}

\begin{table*}[ht!]
\centering
\caption{List of the target sources with the final maps properties.}
\label{tab:targets}
\begin{tabular*}{\textwidth}{ccccccc@{\extracolsep{\fill}}c}
\hline\hline
Target & Redshift & Host galaxy coordinates & Radio morphology & Robust & Resolution & rms noise \\
 & &[J2000] & & & & [$\mu\rm{Jy~beam^{-1}}$]\\
\hline\hline
Double & 0.07844 & $\rm{17^h 13^m 29^s.11}$; $\rm{+64^{\circ} 02' 48".74}$ & FRII & $-0.5$ & $0.30" \times 0.24"$ & 18\\
Original TRG & 0.07981 & $\rm{17^h 12^m 23^s.16}$; $\rm{+64^{\circ} 01' 57".09}$ & FRI-NAT & $-0.5$ & $0.34" \times 0.24"$ & 18\\
Goldfish & 0.08092 & $\rm{17^h 13^m 03^s.78}$; $\rm{+64^{\circ} 07' 01".67}$& FRI-NAT & $-0.5$ & $0.55" \times 0.41"$ & 29\\
Beaver & 0.08295 & $\rm{17^h 13^m 16^s.00}$; $\rm{+63^{\circ} 47' 37".54}$ & FRI-NAT & $-0.5$ & $0.54" \times 0.40"$ & 27\\
Embryo & 0.08001 & $\rm{17^h 15^m 09^s.08}$; $\rm{+64^{\circ} 02' 53".51}$ & FRI-WAT & $0.5$ & $0.55" \times 0.40"$ & 40\\
\hline\hline
\end{tabular*}
\tablefoot{Column 1: target name. Column 2: spectroscopic redshift from~\citet{ahn2012} (for the Double and the Original TRG) and~\citet{alam2015} (for the Goldfish, the Beaver, and the Embryo). Column 3: host galaxy coordinates from~\citet{alam2015} (for the Double) and~\citet{gaia2021} (for all the others). Column 4: radio morphological classification following~\citet{fanaroff1974} (FRI/FRII) and~\citet{1980miley} (NAT/WAT). Column 5: robust parameter used for Briggs weighting during imaging. Column 6: beam size. Column 7: rms noise.}
\end{table*}

\subsection{Self-calibration on target cluster radio galaxies}
\label{sec:fullres_imaging}
With the IS fully calibrated, we imaged the main cluster radio galaxies with LOFAR-VLBI at 144 MHz at sub-arcsecond resolution. For each of the 7 runs, the concatenated datasets produced by the LOFAR-VLBI pipeline were phase-shifted towards five targets and averaged. The selected targets correspond to the main cluster radio galaxies: the Double, the Original TRG, the Goldfish, and the Beaver. First, we applied to all the sources the h5parm, DIE calibration solutions from the in-field calibrator to each observation individually. Given the large distance of the in-field calibrator from the pointing center ($0.8^{\circ}$) and the complex morphology of the targets, we used compact and bright sources closer to the cluster as additional calibrators, to improve the calibration solutions on targets in a way similar to the faceting~\citep{2016vanweeren} or the wide-field direction-dependent calibration~\citep{dejong2024}. The radio source J171259+640931 was used as additional calibrator for the Double and the Goldfish. After doing self-calibration, solving for \texttt{scalarphase} and \texttt{scalarcomplexgain}, and phasing-up the core stations, to reduce the FoV and so the effects from the nearby radio emission from the cluster center, we transferred the calibration solutions to the two targets and proceeded with self-calibration. For the Double, we solved for \texttt{tec}, \texttt{scalarphase}, and \texttt{scalarcomplexgain} avoiding all the baselines below $10,000\lambda$ (corresponding to $\sim 20"$) and phased-up the core stations. For the Goldfish, we solved for \texttt{tecandphase} and \texttt{scalarcomplexgain} and phased-up the superterp stations, avoiding all the baselines below $5,000\lambda$ (corresponding to $\sim 41"$). The Double was also used as additional calibrator for the Original TRG: for the main tailed cluster radio galaxy we solved for \texttt{tec}, \texttt{scalarphase}, and \texttt{scalarcomplexgain} avoiding all the baselines below $10,000\lambda$ and phased-up the core stations. Including shorter baselines for these extended sources, with respect to what was done for the in-field calibrator, was required to add more large-scale signal during calibration. For Beaver and the Embryo instead, we used the radio source J171423+635707 as additional calibrator. Given the dominant low surface brightness extended emission in these sources, self-calibration did not result in any improvement in the image quality: for this reason, we simply applied the solutions from J171423+635707 and imaged the two targets. After self-calibration (where possible), we imaged each source with \texttt{WSClean} using \texttt{wgridder}~\citep{arras2021,ye2022}, combining all the calibrated MS at once, using automatic masking and \texttt{multiscale} deconvolution~\citep{offringa2017}, to better deal with their extended structure.

\section{Morphological analysis of cluster member radio galaxies}
\label{sec:single_targets}
In this Section, we present the final imaging results for the target radio galaxies as well as their morphological detailing. Their main properties are listed in Tab.~\ref{tab:targets}.

\subsection{The Double}
\label{sec:double}
\begin{figure}[h!]
\includegraphics[width=\columnwidth]{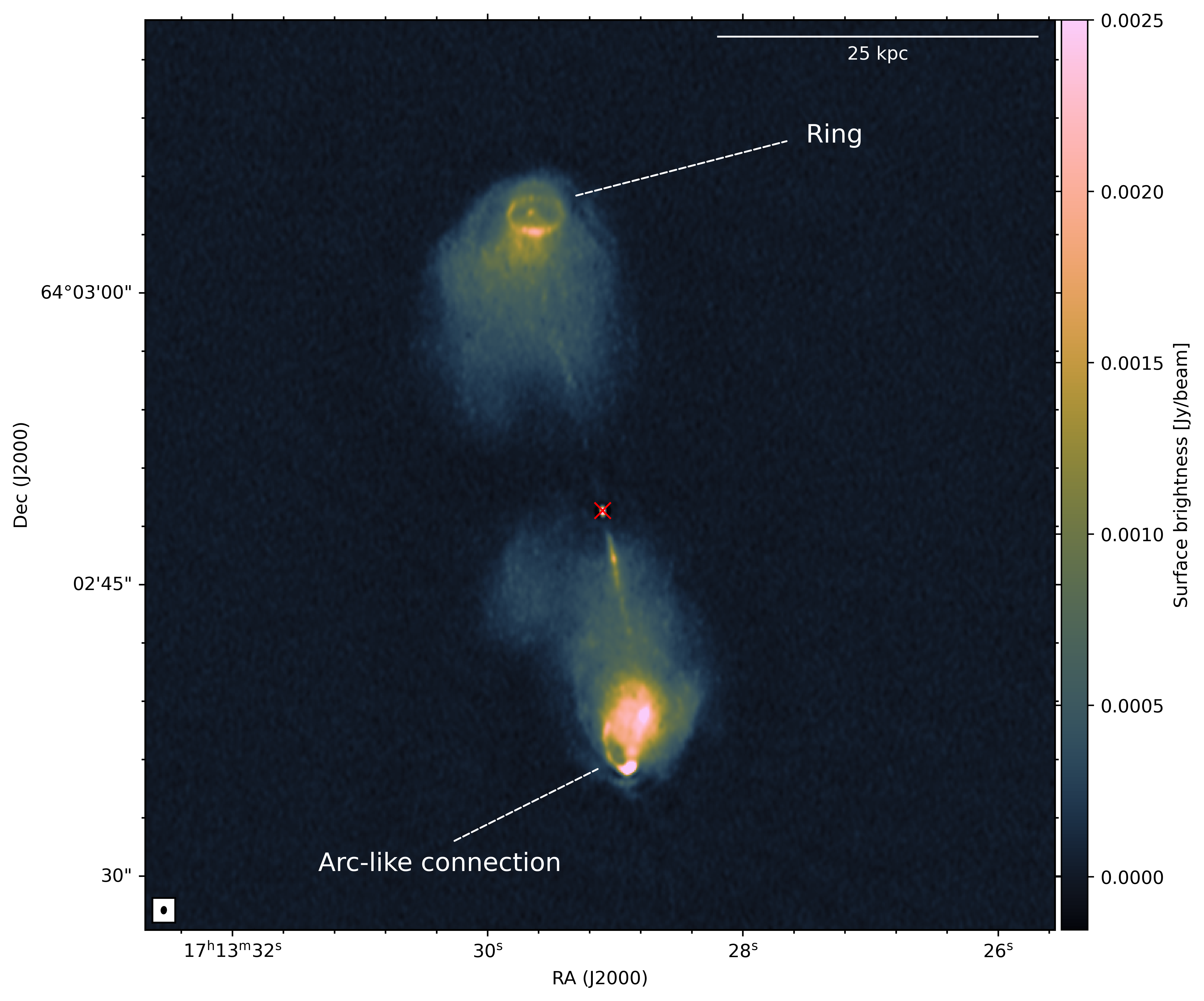}
\caption{The Double at 144 MHz; resolution $0.30" \times 0.24"$, rms noise $18~\rm{\mu Jy~beam^{-1}}$. This image was obtained using Briggs weighting, robust $= -0.5$, and \texttt{multiscale}. The restoring beam size is shown in the bottom-left corner. The red cross identifies the optical position of the host galaxy as listed in Tab.~\ref{tab:targets}.}
\label{fig:double_lofarvlbi}
\end{figure}

The Double, also known as J171329+640249~\citep[][]{miller2002}, is a FRII radio galaxy located, in projection, at about 600 kpc from the cluster center. It exhibits a double-lobed morphology with a total extent of almost 50 kpc ($\approx 35"$). Our new LOFAR-VLBI sub-arcsecond resolution image of the source is shown in Fig.~\ref{fig:double_lofarvlbi}. We resolve the core and the two jets departing from it up to the hotspots. Both lobes show edge-brightening at their tips, possibly resulting from the shocks formed by the interaction with the ICM, and extend for about 21 kpc (north) and 23 kpc (south), measured above the $3\sigma_{\rm{rms}}$. At this detection level, the source has a flux density of $0.95 \pm 0.10$ Jy at 144 MHz. Assuming a spectral index $\alpha = 0.65$~\citep{botteon2020}, the source has a rest-frame luminosity $L_{144} \sim 1.48 \times 10^{25}~\rm{W~Hz^{-1}}$: the classical FRI/II luminosity break is around $L_{150} \sim 10^{26}~\rm{W~Hz^{-1}}$~\citep{fanaroff1974,ledlow1996}, so the Double can be classified as \enquote{FRII-Low}~\citep{mingo2019}. The southern hotspot is shifted with respect to the initial direction of the jet, which ends up into another bright spot in the same lobe: this may suggest the presence of multiple hotspots~\citep{hardcastle1997}. In the southern hotspot there is also a narrow arc-like structure (labeled in Fig.~\ref{fig:double_lofarvlbi}) which, as for other similar objects, could represent a well-collimated backflow~\citep{leahy1997}. The northern hotspot is instead surrounded by a ring-shaped radio brightness enhancement (labeled in Fig.~\ref{fig:double_lofarvlbi}). These rings are unusual, but not unique: they were already observed in other radio galaxies, e.g. in the western hotspot of Hercules A~\citep{dreher1984,timmerman2022}, in 3C 310~\citep{morrison1996}, or in 3C 219~\citep{perley1980}. Several scenarios can explain the formation of the rings~\citep{saxton2002, gizani2003}, involving mainly the shocks and the inner lobes model. In the former, the rings are shocks with circular geometry, caused by backflows in the cocoon, which introduce adiabatic compression and particle acceleration. In the latter, instead, rings arise from material deposited by new jets in the lobe and separated from it by a contact discontinuity. From the sole radio map, disentangling the correct one is non-trivial. The presence of rim-brightening, observed in the ring around the northern hotspot in the Double, favors the shock model because it is the only one that can explain such feature, as also shown in simulations~\citep{meenakshi2023}. Additional information, such as the spectral index, is still required to reliably discriminate between the proposed models or, if necessary, to motivate new ones. Further investigation of all these aspects is beyond the scope of this paper.

\begin{figure*}[h!]
\centering
\includegraphics[width=\textwidth]{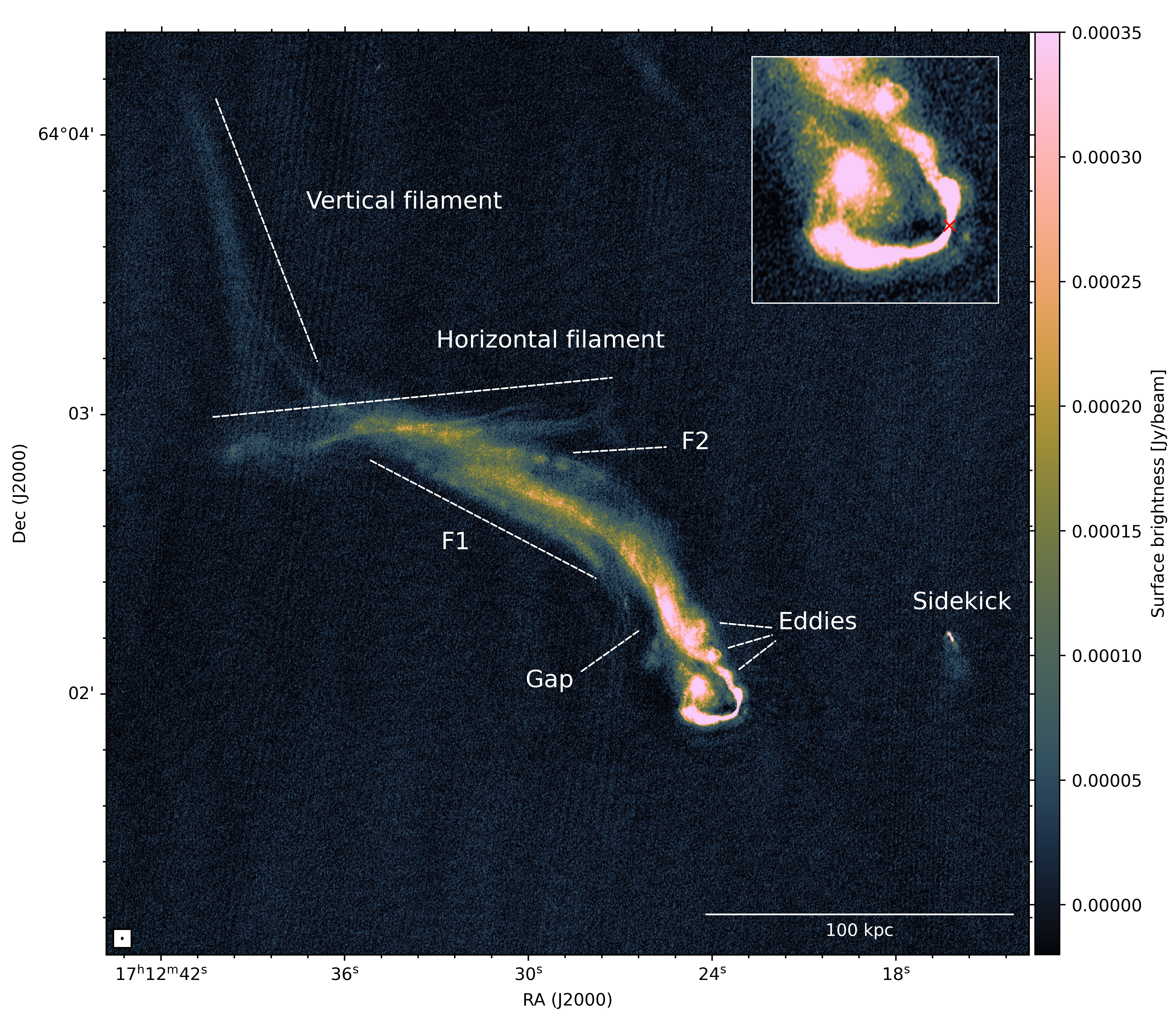}
\caption{The Original Tailed Radio Galaxy (Original TRG) at 144 MHz; resolution $0.34" \times 0.24"$, rms noise $18~\rm{\mu Jy~beam^{-1}}$. This image was obtained using Briggs weighting, robust $= -0.5$, and \texttt{multiscale}. The restoring beam size is shown in the bottom-left corner. In the top-right corner there is a zoom-in ($22~\rm{kpc}$) of a region close to the host galaxy, whose optical position (listed in Tab.~\ref{tab:targets}) is identified with a red cross.}
\label{fig:origtrg_lofarvlbi}
\end{figure*}

\subsection{The Original TRG}
\label{sec:origtrg}
The Original TRG, also known as 7C 1712+6406~\citep[][]{hales2007}, is a NAT radio galaxy, classified as a FRI by~\citet{capetti2017}. Previous studies at higher frequencies~\citep[VLA L-, C-, and X-band][]{govoni2005, govoni2006} reported a source extension of about $2.5'$ with a mean fractional polarization up to $14\%$. Our deep, high-resolution map provided by LOFAR-VLBI is presented in Fig.~\ref{fig:origtrg_lofarvlbi}. We reveal several new sub-structures along the tail, that otherwise would have been considered as a unique, continuous component. First of all, we precisely identify the host galaxy (red cross in Fig.~\ref{fig:origtrg_lofarvlbi}) and the location where the two relativistic jets are ejected. Following the tail extension in the north-east direction, we observe three turbulent eddies at distances between 10 and 30 kpc from the host galaxy, evenly spaced from each other ($\sim 9$ kpc), possibly arising by the instabilities developed in the downstream of the radio galaxy. Interestingly, the entire tail seems to depart only from the northern jet: the southern one, instead, seems to interrupt abruptly, as also observed at higher frequencies by~\citet{govoni2006}. Thanks to the combination of high-sensitivity and resolution we also detect, for the first time, multiple filamentary structures that constitute the tail of the radio galaxy. At 144 MHz we reveal the presence of an extended filament at the lower bound of the main tail, labeled as F1, which follows its shape, just after an emission gap of around 10 kpc from the southern jet termination. This may suggest that some event could have prevented the development of a tail from the southern jet for a fixed amount of time during the host galaxy motion. However, the lack of the southern tail can also result from projection effects, where it can be bent and hidden behind the northern one. The southern jet is connected to a bright emission patch, having flux density $\sim 0.08~\rm{Jy~beam^{-1}}$ (at $10\sigma_{\rm{rms}}$), through two tethers similar to the ones observed in the radio galaxy IC711 in Abell 1314~\citep{vanweeren2021} or in the MysTail in Abell 3266~\citep{rudnick2021}. Together with F1, we detect a horizontal filament at the north-east end of the main tail and a vertical filament that departs from this last one towards north. Between the upper end of the main tail and the west side of the horizontal filament there is another filamentary feature that emerges, labeled as F2 in Fig.~\ref{fig:origtrg_lofarvlbi}. Despite resembling the F1 filament, it is attached directly to the horizontal filament, suggesting a possibly different origin and rather being more connected to what is happening above the horizontal filament. A more quantitative description of the filaments, including their lengths and widths, is presented in Sec.~\ref{sec:filaments_origtrg}.
\par The characteristics of plasma in the main tail and in the filaments cannot be determined solely by morphological analysis, but rather from spectral index analysis. We defer such an analysis to an upcoming paper, where we will present and discuss images of the Original TRG and its filaments obtained at 144, 1260, and 1520 MHz with a resolution of $1.5"$ (De Rubeis et al. \textit{in preparation}). Spectral index information between 49 and 145 MHz from~\citet{botteon2022} ($12.5"$ resolution) on the Original TRG reports nuclear values around $\alpha \sim 0.5-0.6$, with signs of spectral aging along the tail up to $\alpha \sim 2$ corresponding to what we identified as the vertical filament. The combination of low-resolution and low-frequency implies a significant contribution of the diffuse emission from the radio halo in which this radio galaxy is situated: this motivates the follow-up at higher frequency and resolution, to extract the intrinsic spectral properties of the radio galaxy.\\
At almost $50"$ from the Original TRG there is the Sidekick radio galaxy~\citep[$z = 0.071436$,][]{miller2003}, another cluster member radio galaxy for which we resolve its double-jetted structure, which extends for $\sim$ 3.6 kpc and then ends up into a tail that goes south for around 80 kpc.

\subsection{The Goldfish}
\label{sec:goldfish}
\begin{figure}
\includegraphics[width=\columnwidth]{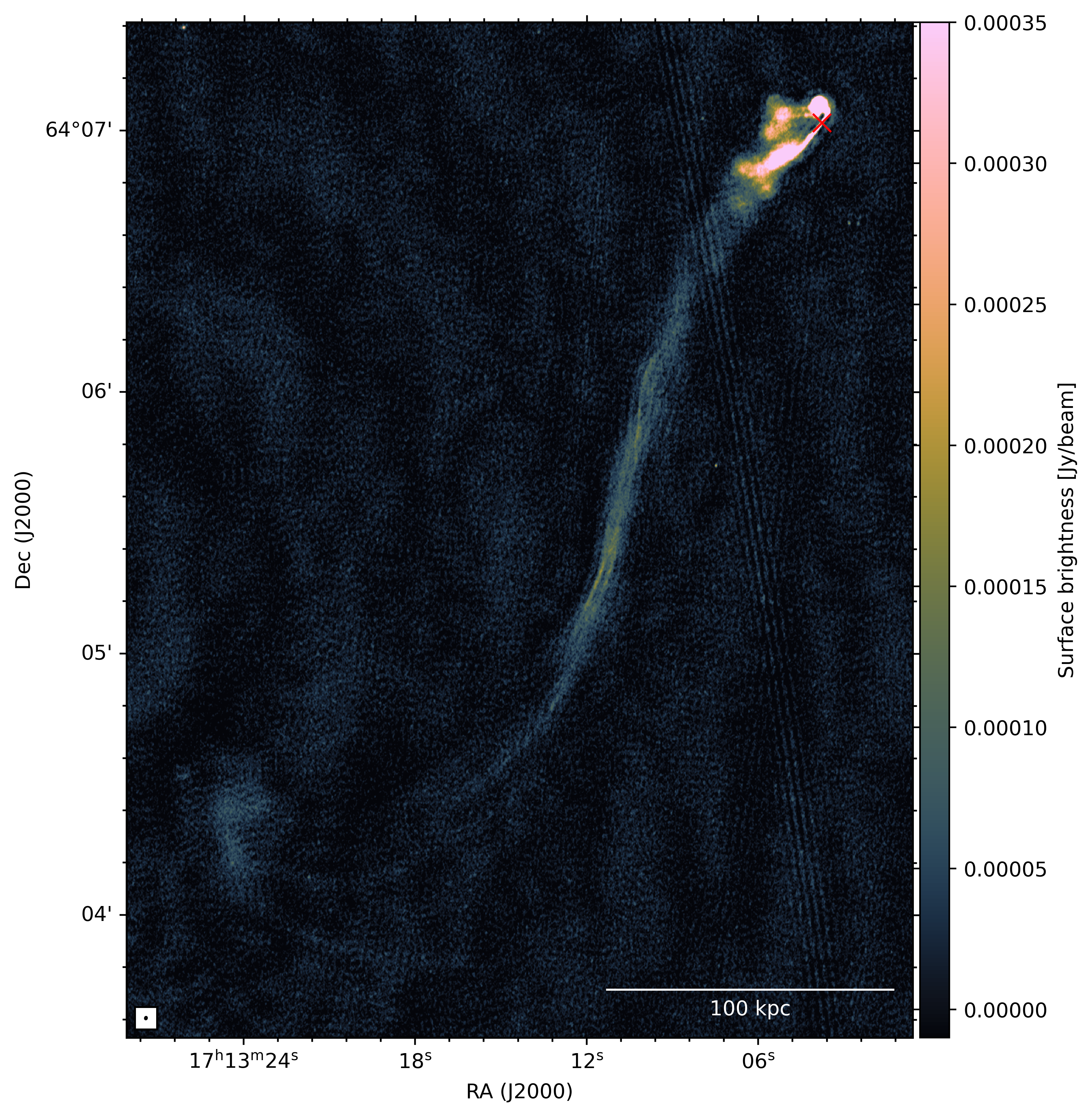}
\caption{The Goldfish at 144 MHz; resolution $0.55" \times 0.41"$, rms noise $29~\rm{\mu Jy~beam^{-1}}$. This image was obtained using Briggs weighting, robust $= -0.5$, and \texttt{multiscale}. The restoring beam size is shown in the bottom-left corner. The red cross identifies the optical position of the host galaxy as listed in Tab.~\ref{tab:targets}.}
\label{fig:goldfish_lofarvlbi}
\end{figure}

The Goldfish, also known as 7C 1713+6407a~\citep[][]{hales2007}, is a NAT radio galaxy classified as FRI. A deep LOFAR image made with Dutch array (Fig.~\ref{fig:lofar_6aseclabel}) showed an extended head and a narrow tail of $\sim 300$ kpc length that ends up mixing with a diffuse feature, named Counter-comma~\citep{botteon2020}, whose origin is still unclear. High-resolution, high-frequency observations (L-, C-, and Ku-band) resolved the head of the radio galaxy, with two jets that depart from the core with a gap of $1.5"$ observed for both of them~\citep{owen1997,ternidegregory2017}. From a deep LOFAR-VLBI map, presented in Fig.~\ref{fig:goldfish_lofarvlbi}), we resolve the core and the region immediately close to the head (within 50 kpc from the host), where the ram pressure effects on the jets can be observed, especially for the northern one, which is deflected by almost 90 degrees as soon as it leaves the core region. At sub-arcsecond resolution we also detect the tail emission up to the connection with the Counter-comma, with the presence of a region of increased surface brightness at around 170 kpc from the core. Interestingly, the tail presents a helical morphology right after the head region, as if the two jets are spiraling around each other before linking back into a single tail that then extends southwards. This wobbling, similar to what happens for the Beaver radio galaxy (Sec.~\ref{sec:beaver}, but on smaller scales), can be due to different reasons related to the motion of the host galaxy or the instabilities developed behind it.

\begin{figure*}[ht!]
\centering
\includegraphics[width=\textwidth]{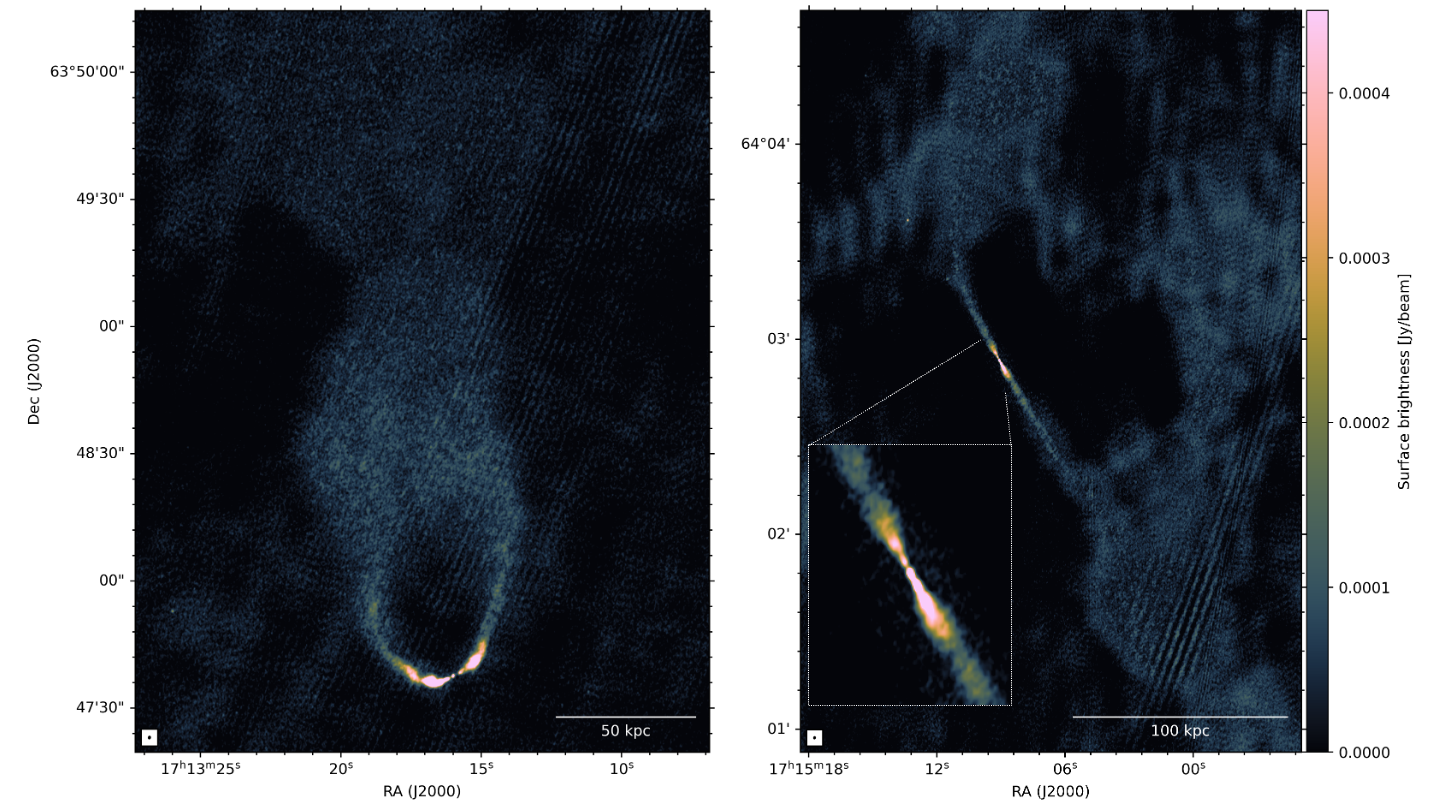}
\caption{Left: the Beaver at 144 MHz, resolution $0.54" \times 0.40"$, rms noise $27~\rm{\mu Jy~beam^{-1}}$. The image was obtained using Briggs weighting, robust $= -0.5$, and \texttt{multiscale}. Right: the Embryo at 144 MHz, resolution $0.55" \times 0.40"$, rms noise $40~\rm{\mu Jy~beam^{-1}}$. The image was obtained using Briggs weighting, robust $= 0.5$, and \texttt{multiscale}. The bottom-left panel shows a zoomed-in view of approximately 32 kpc centered on the source. For both images, the restoring beam size is shown in the bottom-left corner.}
\label{fig:beaver_embryo_lofarvlbi}
\end{figure*}

\subsection{The Beaver}
\label{sec:beaver}
The Beaver, also known as 7C 1712+6352~\citep[][]{hales2007}, is a NAT radio galaxy placed at $\sim 1.5$ Mpc from the cluster center. Low frequency observations made with LOFAR Dutch array at 144 MHz revealed a twisted tail that extends for more than 1 Mpc, before fading into the cluster radio halo (Fig.~\ref{fig:lofar_6aseclabel}). Deep LOFAR-VLBI map is shown in Fig.~\ref{fig:beaver_embryo_lofarvlbi} (left panel). At sub-arcsecond resolution, we resolve mostly the core region and the two jets which depart from it, and also the first, brighter, part of the tail up to the point at which the two jets get close for the first time along the tail, before developing its corkscrew characteristic feature fully depicted in Fig.~\ref{fig:lofar_6aseclabel}. The rest of the tail is not detected because of the lower surface brightness and the time and bandwidth smearing effects, that reduce the intensity far from the pointing center~\citep{bridle1999}.

\subsection{Embryo}
\label{sec:embryo}
The Embryo, also known as GB6 B1714+6405~\citep[][$z=0.08001$]{gregory1996}, is a WAT radio galaxy located, in projection, at about 1.5 Mpc from the cluster center. Its radio emission at 144 MHz extends for $\sim600$ kpc, considering the more diffuse emission generated by ram pressure effects in correspondence of the lobes. The deep, sub-arcsecond resolution map is shown in Fig.~\ref{fig:beaver_embryo_lofarvlbi} (right panel). As for the Beaver, we resolve the core region with the inner jets; the southern jet is brighter than the northern one by a factor of 2, considering the $5\sigma_{\rm{rms}}$ emission within a projected distance of $10.5~\rm{kpc}$ from the core. Moreover, there are multiple bright spots along the northern jet, possibly hinting at the central black hole’s duty cycle. The jets are maintained collimated ($3.5-6$ kpc) before feeling the effects of the ICM and being bent. Following~\citet{laing1999} (Eq. 1), assuming isotropic emission in the source rest-frame and a spectral index $\alpha=0.5$~\citep[from][]{botteon2020}, we can constrain the angle to the line of sight to be $\theta \leq 60^{\circ}$ and the jet velocity to be $\geq 0.5 c$. The morphology of the Embryo suggests possible hints for the kinematics of the host galaxy within the ICM: the small bending of the northern jet can be caused by the motion of the host galaxy towards the north-east direction, with the southern jet which is not affected by such dynamics.

\begin{figure}[ht!]
\centering
\includegraphics[width=\columnwidth]{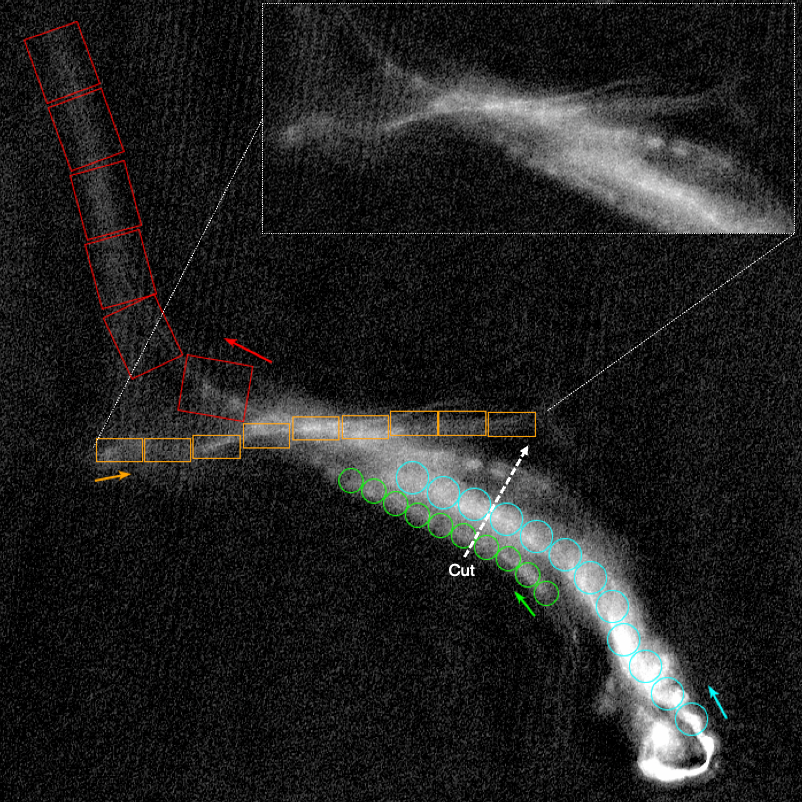}
\caption{Same LOFAR-VLBI map for the Original TRG shown as in Fig.~\ref{fig:origtrg_lofarvlbi} highlighting the regions used for determining the main morphological features associated to the radio galaxy using different colors. Arrow represent the direction of the transversal cut used to retrieve the filaments' width depicted in the bottom panel of Fig.~\ref{fig:flux_profiles}. Top right corner: details of the horizontal filament to highlight the main morphological patterns coexisting in this region.}
\label{fig:origtrg_regions}
\end{figure}
\begin{figure}[h!]
\centering
\includegraphics[width=\columnwidth]{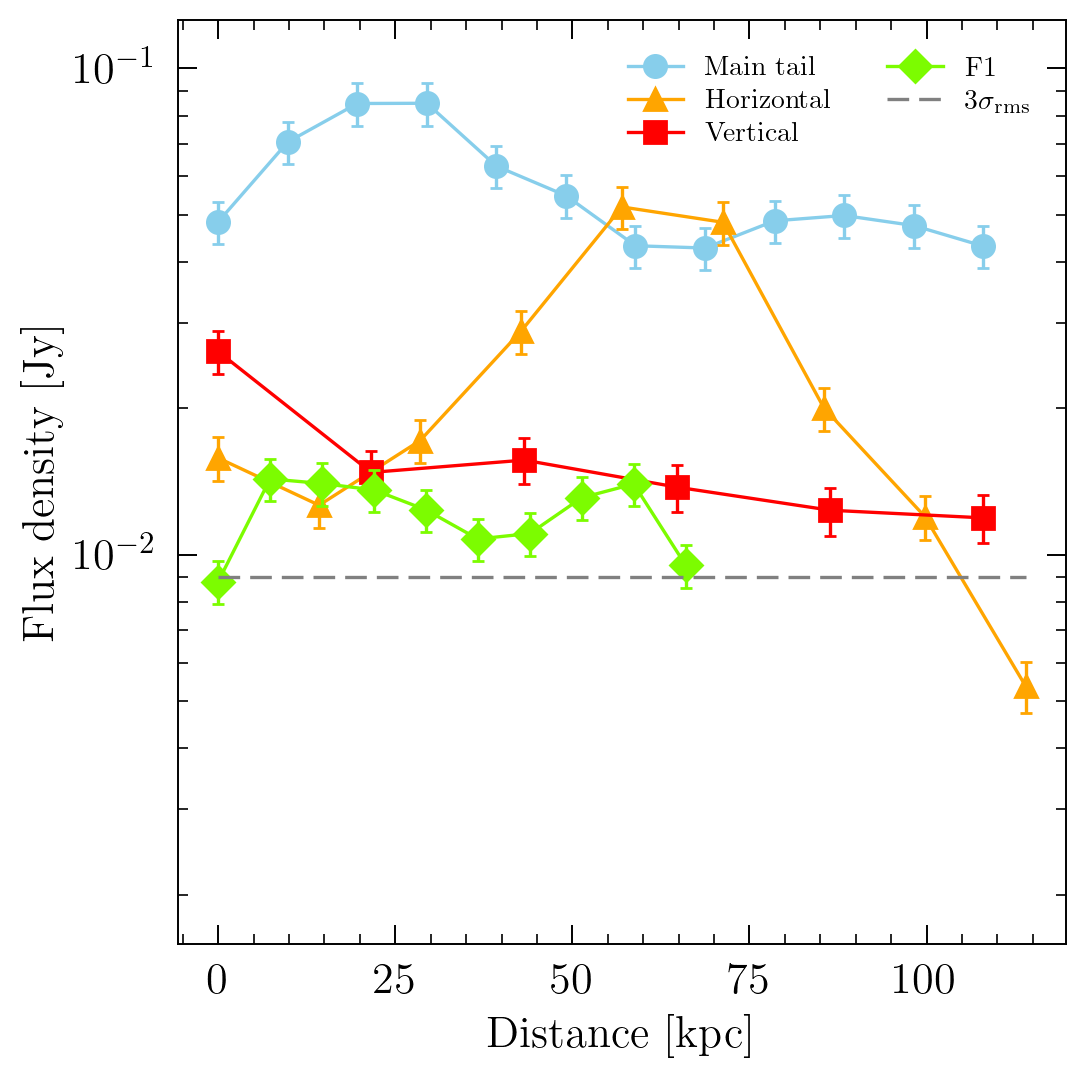}
\includegraphics[width=\columnwidth]{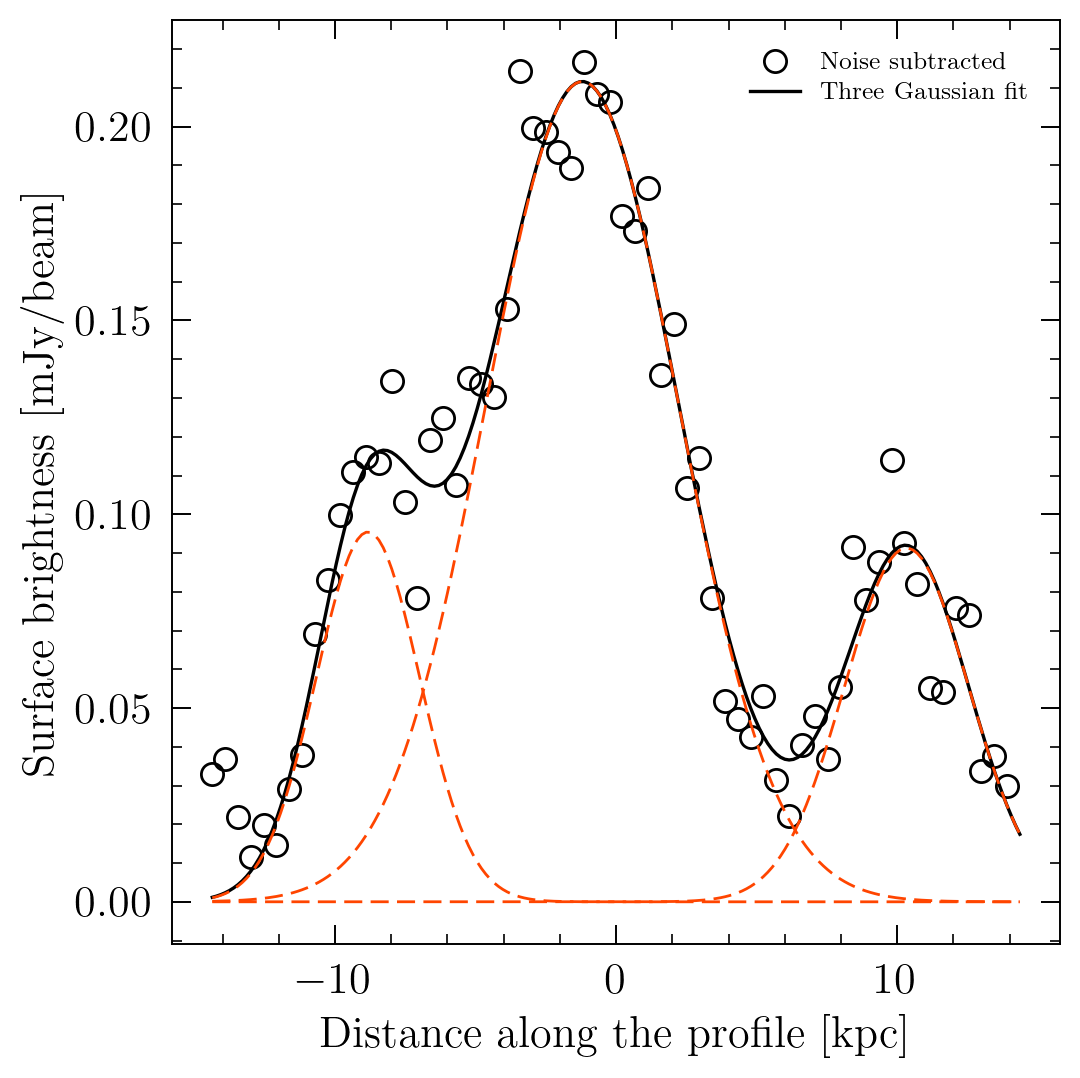}
\caption{Spatial trends along the Original TRG and its filaments from regions highlighted in Fig.~\ref{fig:origtrg_regions}. Top: flux density along the main features of the Original TRG, highlighted in different colors. Error bars refer to the errors on flux densities, calculated assuming a $10\%$ flux uncertainty~\citep[as for the LoTSS,][]{shimwell2022}. Colored arrows indicate the direction followed by the spatial trends. Bottom: profile across the Original TRG following the cut depicted in Fig.~\ref{fig:origtrg_regions}. The noise-subtracted data are shown as open circles. The black solid line shows the total fit using three-Gaussian fit. The three Gaussians (individually shown with red dashed lines) correspond, from left to right, to F1, the main tail, and F2, respectively.}
\label{fig:flux_profiles}
\end{figure}

\section{Characterization of the filaments in the Original TRG}
\label{sec:filaments_origtrg}
In this Section, we detail and discuss the properties related to the morphology and the emissivity of the filaments detected in the Original TRG, based on what we observed at 144 MHz from LOFAR-VLBI. We distinguished several morphological features presented in Sec.~\ref{sec:origtrg} and highlighted in Fig.~\ref{fig:origtrg_lofarvlbi}. Projected lengths can be directly measured from the map. The main tail, which develops from the northern jet, extends for $\sim$ 140 kpc before linking with the horizontal filament. The horizontal filament extends for about 110 kpc on the east-west direction, with a forking on the west side. The vertical filament's length is $\sim 105$ kpc. The F1 filament has length of $\sim$ 83 kpc, measured from the gap to the horizontal filaments (where it ends up); F2 is instead more complicated to isolate with respect to the other features, given the tight connection with the horizontal filament and the upper end of the main tail: in this case, spectral index information will be helpful in determining the properties of its electron population, and so its origin. Overall, considering all three components (main tail, horizontal and vertical filament) the Original TRG complex extends for at least $\sim 3'$ ($\sim 270$ kpc).
\par To measure the width of the filaments and the main tail, we traced multiple transversal cuts (as the one showed in Fig.~\ref{fig:origtrg_regions}, dashed white arrow) across the Original TRG, obtaining a spatial surface brightness profile for each cut. The rms noise of the image was quadrature-subtracted to the brightness values and the resulting profiles were fitted with single/multiple Gaussian profiles. For the fit, we used the non-linear least squares method. The number of Gaussians of the model function was chosen by visual inspection of the surface brightness profiles along the cuts. The amplitude, standard deviation, and center of each Gaussian were allowed to vary: as initial guess, we used values of $1.0$ for the amplitude and the standard deviation, while for the center we used equidistant values (depending on the number of Gaussians) distributed within the length of the cut. The beam-deconvolved full-width at half-maximum of the fitting Gaussian is then considered to be the width of the corresponding feature. An example of three-Gaussian fit for the central part of the Original TRG, which includes F1, main tail, and F2, is shown in the bottom panel of Fig.~\ref{fig:flux_profiles}. The main tail has widths that start from $1.8$ kpc (at the point where it starts developing from the northern jet) to $3.5-4.7$ kpc between the eddies. Moving upward, the width ranges between 8 and 13 kpc: then, it mixes up with the horizontal filament. The F1 filament has width ranging between 3 and 4.2 kpc (from the gap upward); F2, instead, exhibits a width of 5 kpc. The horizontal filament, when observed at such high-resolution, is resolved into multiple substructures rather than appearing as a single monolithic structure, as highlighted in the top panel of Fig.~\ref{fig:origtrg_regions}. We measured its width in the central brighter part, where there is enough signal-to-noise, to be $\sim 8-10$ kpc, even if there it is difficult to distinguish between the main tail, the horizontal filament and F2 as they are co-spatial. At the edges the detection is at the noise level. The vertical filament has almost constant width of about 4.5 kpc, with a forking at the bottom where it connects in two different parts to the horizontal one.
\par In Fig.~\ref{fig:flux_profiles} we show also the flux density profiles along the main feature highlighted for the Original TRG in Fig.~\ref{fig:origtrg_regions}: the main tail (in cyan), F1 (in green), the horizontal filament (in orange) and the vertical one (in red). The brightness increase in the center of the horizontal filament is due, at least in part, to the overlap with the upper end of the main tail and (apparently) the F2 filament. High-resolution spectral index studies are crucial to disentangle the different emitting components and reconstruct their origin in this part. At high angular resolution, filaments also show their non-thermal emission in multiple patches: this may indicate that their formation process does not act homogeneously along the filaments' extension, but rather have some local contribution which enhances the radio emission.
\par Having thin and long non-thermal filaments, with the horizontal and vertical ones that also show a particularly straight shape, may suggest a common formation scenario with the E-fils observed by~\citet{rudnick2022} in the radio galaxy 3C40B in Abell 194. In that case, the preferred formation scenario relied on the amplification of the magnetic field through shear stretching of magnetic field lines: the observed filaments would then be made by smaller bundles of illuminated and aligned fibers, held together by tension along the field lines. The thickness of the fibers should reflect the transverse resistivity scales, while their length reflects the driving scales of the local turbulent flow~\citep{2018vazza}. For A2255, given the presence of turbulence associated to the merging state of the cluster, this can reach the cluster scale. 
From the filaments' lengths we can evaluate its dynamical lifetime, which is essentially the cascading time of the turbulent eddy on scale $L$. Assuming a turbulence Mach number of $\mathcal{M}=1/2$ and a characteristic ICM sound speed $c_s \sim 10^{3}~\rm{km/s}$~\citep{porter2015}, we end up with
\begin{equation}
    \label{eq:dynamic_age}
    \tau_{\rm{dyn}} \sim \frac{L}{\mathcal{M} \cdot c_s} \sim \frac{L}{0.5}~\rm{Myr}.
\end{equation}
Our filaments have lengths ranging between 80-110 kpc, resulting in $\tau_{\rm{dyn}} \sim 160-220$ Myr: this is comparable to the synchrotron radiative lifetimes ($10^{7}-10^8$ yr), meaning that electrons have sufficient time to emit all their energy via synchrotron radiation before the filament gets dissipated by the turbulence. Spectral index information along the filaments will enable us to constrain particle acceleration, if present, as well as the spatial diffusion coefficient required to observe radio emitting electrons on such scales. Stretching of the magnetic field lines implies magnetic field amplification by a factor proportional to $\rho v^2/l$, where $\rho$ is the density of the turbulent ICM, with characteristic velocity $v$ on the driving scale $l$. This causes enhancement in the magnetic field pressure $P_{B}=B^2/8\pi$, which becomes compatible with the thermal pressure, reducing the plasma beta of the filaments $\beta_F = P_{gas}/P_B$. We expect the filaments, given also their straightness, to have very low plasma beta (approaching unity), being high-magnetic pressure regions. In support of this scenario, some cases have shown an overlap between the presence of non-thermal radio filaments and the absence of thermal, X-ray emitting ICM~\citep{rudnick2022}.

\section{Conclusions}
\label{sec:conclusion}
We presented LOFAR-VLBI observations at sub-arcsecond resolution of the brightest cluster radio galaxies in the merging cluster Abell 2255, obtained using 56 hours of observations. These images represent, to date, the deepest ones ever obtained using LOFAR-VLBI for a galaxy cluster. The unique resolution and sensitivity capabilities of LOFAR IS showed unprecedented structures connected to the radio galaxies, particularly to the Original TRG where several filaments have been discovered around the tail and attached to its upper end. Despite being already observed in other galaxy clusters/groups, especially in presence of a turbulent ambient medium, this represent the first time that these objects have been observed at such high-resolution, proving the potential of LOFAR-VLBI observations for these elongated, non-thermal filaments. They have lengths ranging from 80 to 110 kpc, with varying widths along their extension ($3-10$ kpc), as other cases observed in other clusters~\citep[such as in Abell 194, ][]{rudnick2022}. Being thin and straight, there is the possibility that, as the ones in A194, they present very low plasma beta due to the amplification of magnetic field caused by shear stretching of field lines. One peculiarity observed is the patchiness of the radio emission along the filaments, which may suggest further hints about their formation scenario. To constrain the nature and the origin of these filaments, we will present in an upcoming paper observations at higher frequencies of the Original TRG using uGMRT (Band 5) and VLA (L-band). Providing a resolution of $1.5"$, these images will allow high-resolution spectral index analysis of the filaments to constrain their radio spectral properties and disentangle multiple spectral components that are co-spatial in the Original TRG from lower-resolution studies. Together with this, we will present also additional filaments detected with LOFAR-VLBI at $1.5"$ resolution on larger scales extending for $\sim 250$ kpc, east of the Original TRG, originally known as Trail~\citep{botteon2020}. VLA data will also allow polarization analysis, which has been observed in similar structures up to $40-50\%$~\citep{rudnick2022}. 

\begin{acknowledgements}
EDR and CG are supported by the Fondazione ICSC, Spoke 3 Astrophysics and Cosmos Observations. National Recovery and Resilience Plan (Piano Nazionale di Ripresa e Resilienza, PNRR) Project ID CN\_00000013 \enquote{Italian Research Center for High-Performance Computing, Big Data and Quantum Computing} funded by MUR Missione 4 Componente 2 Investimento 1.4: Potenziamento strutture di ricerca e creazione di \enquote{campioni nazionali di R\&S (M4C2-19)} - Next Generation EU (NGEU). MB acknowledges support from INAF under the Large Grant 2022 funding scheme (project \enquote{MeerKAT and LOFAR Team up: a Unique Radio Window on Galaxy/AGN co-Evolution}). JMGHJdJ acknowledges support from project CORTEX (NWA.1160.18.316) of research programme NWA-ORC, which is (partly) financed by the Dutch Research Council (NWO), and support from the OSCARS project, which has received funding from the European Commission’s Horizon Europe Research and Innovation programme under grant agreement No. 101129751.
This manuscript is based on data obtained with the International LOFAR Telescope (ILT). LOFAR~\citep{vanhaarlem2013} is the Low Frequency Array designed and constructed by ASTRON. It has observing, data processing, and data storage facilities in several countries, which are owned by various parties (each with their own funding sources), and which are collectively operated by the ILT foundation under a joint scientific policy. The ILT resources have benefited from the following recent major funding sources: CNRS-INSU, Observatoire de Paris and Université d’Orléans, France; BMBF, MIWF-NRW, MPG, Germany; Science Foundation Ireland (SFI), Department of Business, Enterprise and Innovation (DBEI), Ireland; NWO, The Netherlands; The Science and Technology Facilities Council, UK; Ministry of Science and Higher Education, Poland; The Istituto Nazionale di Astrofisica (INAF), Italy. This research made use of the LOFAR-IT computing infrastructure supported and operated by INAF, including the resources within the PLEIADI special \enquote{LOFAR} project by USC-C of INAF, and by the Physics Dept. of Turin University (under the agreement with Consorzio Interuniversitario per la Fisica Spaziale) at the C3S Supercomputing Centre, Italy. This research made use of Matplotlib~\citep{hunter2007}, APLpy~\citep[an open-source plotting package for Python,][]{robitaille2012}, Astropy~\citep[a community-developed core Python package and an ecosystem of tools and resources for astronomy,][]{astropy2022}.
\end{acknowledgements}

\bibliographystyle{aa}
\bibliography{bibl}

\end{document}